\documentclass[conference]{IEEEtran}
\ifCLASSINFOpdf
\else
\fi

\usepackage{tikz} 
\usepackage{amsmath}
\usepackage{amssymb}
\usepackage{algorithm}
\usepackage{algpseudocode}

\begin{document}
%
\title{Binary Compressive Sensing via \\ Smoothed $\ell_0$ Gradient Descent}

\author{\IEEEauthorblockN{Tianlin Liu}
\IEEEauthorblockA{Department of Computer Science and Electrical Engineering\\
Jacobs University Bremen\\
28759 Bremen, Germany\\
Email: t.liu@jacobs-university.de}
\and
\IEEEauthorblockN{Dae Gwan Lee}
\IEEEauthorblockA{Mathematisch-Geographische Fakult\"at\\
Katholische Universit\"at Eichst\"att-Ingolstadt\\
85071 Eichst{\"a}tt, Germany\\
Email: daegwans@gmail.com}}


%


\maketitle

\begin{abstract}
We present a Compressive Sensing algorithm for reconstructing binary signals from its linear measurements.
The proposed algorithm minimizes a non-convex cost function expressed as a weighted sum of smoothed $\ell_0$ norms which takes into account the binariness of signals.
We show that for binary signals the proposed algorithm outperforms other existing algorithms in recovery rate while requiring a short run time.
\end{abstract}

\section{Introduction}
Compressive Sensing (CS) is a method in signal processing which aims to reconstruct signals from a relatively small number of measurements. 
It has been shown that sparse signals can be reconstructed with a sampling rate far less than the Nyquist rate by exploiting the sparsity \cite{Donoho06}.

In this paper, we focus on Binary Compressive Sensing (BCS) which restricts the signals of interest to binary $\{ 0, 1 \}$-valued signals, which are widely used in engineering applications, such as fault detection \cite{Bickson11}, single-pixel image reconstruction~\cite{Duarte08}, and digital communications~\cite{Wu11}.
Related works are as follows: Nakarmi and Rahnavard \cite{Nakarmi12} designed a sensing matrix tailored for binary signal reconstruction. Wang et al.~\cite{Wang13} combined $\ell_1$ norm with $\ell_\infty$ norm to reconstruct sparse binary signals. Nagahara \cite{Nagahara15} exploited the sum of weighted $\ell_1$ norms to effectively reconstruct signals whose entries are integer-valued and, in particular, binary signals and bitonal images. Keiper et al.~\cite{Keiper17} analyzed the phase transition of binary Basis Pursuit.

We note that most of the previous work on BCS are based on convex optimization. Indeed, convex optimization based algorithms allow performance guarantee via rich mathematical tools. However, they are found to be notoriously slow in large-scale applications compared to greedy methods such as the Orthogonal Matching Pursuit (OMP) \cite{Donoho08}. On the other hand, greedy methods like OMP are fast but often have a worse recovery rate than convex optimization methods. In this work, we propose a fast BCS algorithm with a high recovery rate. Taking the binariness of signals into account, our algorithm is a gradient descent method based on the smoothed $\ell_0$ norm \cite{Mohimani09}. Through numerical experiments, we show that the proposed algorithm compares favorably against previously proposed CS and BCS algorithms in terms of recovery rate and speed.

The rest of the paper is organized as follows. We give a short review on CS/BCS algorithms in Section \ref{sec:BCS} and present our algorithm in Section \ref{sec:BSSL0}. In Section \ref{sec:Experiments}, we present experimental results which compare the performance of the proposed algorithm with other algorithms. We conclude this paper with some remarks in Section \ref{sec:conclusion}.

\subsection*{Notations:}

For a vector $\mathbf{v} = (v_1, \cdots, v_N)^\top$ and $1 \leq p \leq \infty$, the $\ell_p$ norm of $\mathbf{v}$ is denoted by $\| \mathbf{v} \|_p$. The number of non-zero entries in $\mathbf{v}$ is denoted by $\|\mathbf{v}\|_0$. The probability of an event $E$ is denoted by $\mathbb{P}(E)$. Let $[N]=\{1,\cdots,N\}$ for $N \in \mathbb{N}$. We denote by $\textbf{1}_N$ the $N$-dimensional vector with all entries equal to $1$.

\section{Binary Compressive Sensing (BCS)}
\label{sec:BCS}

In the standard CS scheme, one aims to recover a sparse signal from its linear measurements. The constraints posed by the measurements can be formulated as
\begin{equation}
\Phi \, \mathbf{z} = \mathbf{y},
\quad \mathbf{z} \in \mathbb{R}^N,
 \label{eq:measurement}
\end{equation}
where $\Phi \in \mathbb{R}^{m \times N}$, $m \ll N$, is the measurement matrix 
and $\mathbf{y} = \Phi \mathbf{x} \in \mathbb{R}^m$ is the measurement of a \emph{sparse} signal $\mathbf{x} \in \mathbb{R}^N$. CS algorithms exploit the fact that $\mathbf{x}$ is sparse and seek a sparse solution $\mathbf{z}$ satisfying (\ref{eq:measurement}).

The BCS scheme considers binary signals for $\mathbf{x}$. Note that a binary signal $\mathbf{x}$ is sparse if and only if its complementary binary signal $\widetilde{\mathbf{x}} := \textbf{1}_N - \mathbf{x}$ is dense, i.e., is almost fully supported. As the measurement matrix $\Phi$ is known, the equation (\ref{eq:measurement}) converts equivalently to
\begin{equation}
\Phi \widetilde{\mathbf{z}} = \widetilde{\mathbf{y}},
 \label{eq:measurement_dense}
\end{equation}
where $\widetilde{\mathbf{z}} := \textbf{1}_N - \mathbf{z}$ and $\widetilde{\mathbf{y}} := \Phi \widetilde{\mathbf{x}} = \Phi \textbf{1}_N - \mathbf{y}$. This shows that reconstructing a sparse signal $\mathbf{z}$ under the constraint (\ref{eq:measurement}) is equivalent to reconstructing a dense signal $\widetilde{\mathbf{z}}$ under the constraint (\ref{eq:measurement_dense}). For this reason, in contrast to the case of generic signals, binary signals that are dense can be recovered as well as those that are sparse.

Two types of models for binary signals have been considered in the literature (e.g., \cite{Donoho10,Wang13,Nagahara15}):
(i) $\mathbf{x}$ is a deterministic vector which is binary and sparse, i.e., most of its entries are $0$ and only few are $1$; (ii) $\mathbf{x}$ is a random vector whose entries are independent and identically distributed (i.i.d.) with probability distribution $\mathbb{P}(x_j = 1) = p$ for some fixed $0 \leq p \leq 1$. If $p$ is small, a realization of $\mathbf{x}$ is likely a sparse binary signal.

In this work, we shall consider the second model which can accommodate dense binary signals as well as sparse binary signals.

Below we give a short review of CS/BCS methods that are related to our work.

\subsection{$\ell_0$ minimization (L0)}

A naive approach to finding sparse solutions is the $\ell_0$ minimization,
\begin{equation}
\begin{aligned}
& \underset{\mathbf{z} \in \mathbb{R}^N}{\text{min}}
 & &  \|\mathbf{z}\|_0 & & \text{subject to} & & \Phi \mathbf{z} = \mathbf{y}.
\end{aligned}
\tag{$P_0$}
\label{eq:P_0}
\end{equation}
This method works generally for continuous-valued signals that are sparse, i.e., signals whose entries are mostly zero. However, solving the $\ell_0$ minimization requires a combinatorial search and is therefore NP-hard \cite{Natarajan95}.

\subsection{Smoothed $\ell_0$ minimization (SL0)}

Smoothed $\ell_0$ minimization (SL0) \cite{Mohimani09} replaces the $\ell_0$ norm in (\ref{eq:P_0}) with a non-convex relaxation:
\begin{equation*}
\begin{aligned}
& \underset{\mathbf{z} \in {\mathbb{R}}^N}{\text{min}}
 & &  \sum_{i = 1}^N \left ( 1 - \exp \left (\frac{- z_i^2}{2 \sigma^2} \right ) \right )  && \text{subject to} & & \Phi \mathbf{z} = \mathbf{y}.
\end{aligned}
\label{eq:SL0}
\end{equation*}
This is motivated by the observation 
\[ \lim_{\sigma \to 0} \exp \left( \frac{-t^2}{2 \sigma^2} \right) =  \begin{cases}
      1 & \text{if~} t = 0 \\
      0 & \text{if~}  t \neq 0,
   \end{cases}
\]
which implies that for any $\mathbf{z} = (z_1, \dots, z_N )^\top \in \mathbb{R}^N$,
\begin{equation}
\lim_{\sigma \to 0}  \sum_{i = 1}^N \left( 1 - \exp \left( \frac{-z_i^2}{2 \sigma^2} \right) \right) = \| \mathbf{z} \|_0 .
\label{eq:SL0convergence}
\end{equation}
Noticing that $\mathbf{z} \mapsto \sum_{i = 1}^N \big( 1 - \exp \big( \frac{-z_i^2}{2 \sigma^2} \big) \big)$ is a smooth function for any fixed $\sigma > 0$, Mohimani et al.~\cite{Mohimani09} proposed an algorithm based on the gradient descent method. The algorithm iteratively obtains an approximate solution by decreasing $\sigma$.

Mohammadi et al.~\cite{Mohammadi14} adapted the SL0 algorithm particularly to non-negative signals. Their algorithm, called the Constrained Smoothed $\ell_0$ method (CSL0), incorporates the non-negativity constraints by introducing some weight functions into the cost function. Empirically, CSL0 shows better performance than SL0 in the reconstruction of non-negative signals.


\subsection{Basis Pursuit (BP)}

A well-known and by now standard relaxation of (\ref{eq:P_0}) is the $\ell_1$-minimization, also known as the \emph{Basis Pursuit} (BP) \cite{Chen01}:
\begin{equation}
\begin{aligned}
& \underset{\mathbf{z} \in \mathbb{R}^N}{\text{min}}
 & &  \|\mathbf{z}\|_1 && \text{subject to} & & \Phi \mathbf{z} = \mathbf{y}.
\end{aligned}
\tag{$P_1$}
\label{eq:P_q}
\end{equation}
Similar to (\ref{eq:P_0}), this method works generally for continuous-valued signals $\mathbf{x} \in \mathbb{R}^N$ that are sparse.

\subsection{Boxed Basis Pursuit (Boxed BP)}

Donoho et al.~\cite{Donoho10} proposed the Boxed Basis Pursuit (Boxed BP) for the reconstruction of \emph{k-simple bounded signals}:
\begin{equation*}
\begin{aligned}
& \underset{\mathbf{z} \in [0,1]^N}{\text{min}}
 & & \|\mathbf{z}\|_1 && \text{subject to} & & \Phi \mathbf{z} = \mathbf{y}.
\end{aligned}
\end{equation*}
The intuition behind Boxed BP is straightforward: the $\ell_1$ norm minimization promotes sparsity of the solution while the restriction $\mathbf{z} \in [0,1]^N$ reduces the set of feasible solutions. Recently, Keiper et al.~\cite{Keiper17} analyzed the performance of Boxed BP for reconstructing binary signals.

\subsection{Sum of Norms (SN)}
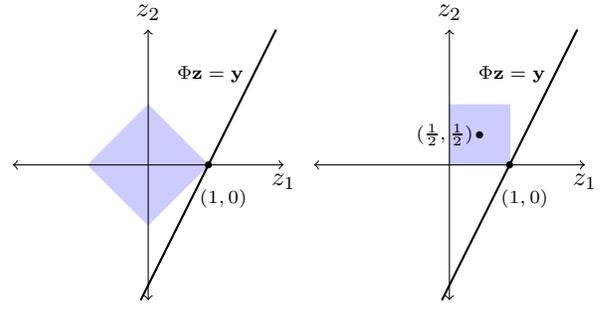
\begin{figure}[t]
  \centering
\begin{tikzpicture}[scale=0.2]
\def\L{9}
\def\Lminusone{8}
\filldraw [fill=blue!20, draw=blue!20] (4,0) -- (0,4) -- (-4,0) -- (0,-4) -- (4,0);
\draw[<->] (-\L,0) -- (\L,0) node[below] {$z_1$};
\draw[<->] (0,-\L) -- (0,\L) node[above] {$z_2$};
\draw[-, thick] (-0.5,-9) -- (8.5,9);
\draw (4,0) circle (2mm) [fill=black];
\node[below] at (5,-1) {\scriptsize $(1,0)$};
\node[left] at (7,6) {\scriptsize $\Phi \mathbf{z} = \mathbf{y}$};
\end{tikzpicture}
\begin{tikzpicture}[scale=0.2]
\def\L{9}
\def\Lminusone{8}
\filldraw [fill=blue!20, draw=blue!20] (4,0) -- (4,4) -- (0,4) -- (0,0) -- (4,0);
\draw[<->] (-\L,0) -- (\L,0) node[below] {$z_1$};
\draw[<->] (0,-\L) -- (0,\L) node[above] {$z_2$};
\draw[-, thick] (-0.5,-9) -- (8.5,9);
\draw (4,0) circle (2mm) [fill=black];
\node[below] at (5,-1) {\scriptsize $(1,0)$};
\draw (2,2) circle (2mm) [fill=black];
\node[left] at (2.2,2) {\scriptsize $(\frac{1}{2},\frac{1}{2})$};
\node[left] at (7,6) {\scriptsize $\Phi \mathbf{z} = \mathbf{y}$};
\end{tikzpicture}
  \caption{Left: the minimization of $\|\mathbf{z}\|_1$ finds sparse solutions. Right: the minimization of $\| \mathbf{z} - \frac{1}{2} \cdot \textbf{1}_N \|_{\infty}$ forces the entries $|z_i - \frac{1}{2}|$ to be small and of equal magnitude.}
  \label{fig:ell1-ellinfty}
\end{figure}

Wang et al. \cite{Wang13} introduced the following optimization problem which combines the $\ell_1$ and $\ell_{\infty}$ norms:
\begin{equation*}
\begin{aligned}
& \underset{\mathbf{z} \in \mathbb{R}^N}{\text{min}}
 & &  \|\mathbf{z}\|_1 + \lambda \, \| \mathbf{z} - \tfrac{1}{2} \, \textbf{1}_N \|_{\infty} && \text{subject to} & & \Phi \mathbf{z} = \mathbf{y}.
\end{aligned}
\end{equation*}
Minimizing $\|\mathbf{z}\|_1$ promotes sparsity of $\mathbf{z}$ while minimizing $\| \mathbf{z} - \frac{1}{2} \, \textbf{1}_N \|_{\infty}$ forces the entries $|z_i - \frac{1}{2}|$ to be small and of equal magnitude (see Fig.~\ref{fig:ell1-ellinfty}).
The two terms are balanced by a tuning parameter $\lambda >0$.

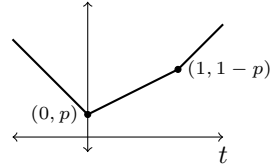
\begin{figure}[t]
  \centering
\begin{tikzpicture}[scale=0.2]
\def\L{9}
\draw[<->] (-5,0) -- (\L,0) node[below] {$t$};
\draw[<->] (0,-1) -- (0,\L);
\draw[-, thick] (-5,6.5) -- (0,1.5);
\draw[-, thick] (0,1.5) -- (6,4.5);
\draw[-, thick] (6,4.5) -- (9,7.5);
\draw (6,4.5) circle (2mm) [fill=black];
\draw (0,1.5) circle (2mm) [fill=black];
\node[left] at (0,1.5) {\scriptsize $(0,p)$};
\node[right] at (6,4.5) {\scriptsize $(1,1-p)$};
\end{tikzpicture}
\caption{The function $f$ given in (\ref{eq:ftn_f}).}
  \label{fig:SAV}
\end{figure}

\subsection{Sum of Absolute Values (SAV)}

Nagahara \cite{Nagahara15} proposed the following method for reconstruction of discrete signals whose entries are chosen independently from a set of finite alphabets $\alpha = \{\alpha_1, \alpha_2, \dots, \alpha_L\}$ with a priori known probability distribution. In the special case $\alpha = \{0, 1\}$ of binary signals, SAV is formulated as,
\begin{equation*}
\begin{aligned}
& \underset{\mathbf{z} \in {\mathbb{R}}^N}{\text{min}}
 & &  (1- p) \, \| \mathbf{z} \|_1 + p \, \| \mathbf{z} -  \textbf{1}_N \|_1 && \text{subject to} & & \Phi \mathbf{z} = \mathbf{y},
\end{aligned}
\end{equation*}
where $p = \mathbb{P}(x_j = 1)$, $j \in [N]$, is the probability distribution of the entries of $\mathbf{x}$.
If $p \approx 0$, i.e., if $\mathbf{x}$ is sparse, then $(1- p) \, \| \mathbf{z} \|_1 + p \, \| \mathbf{z} -  \textbf{1}_N \|_1 \approx \| \mathbf{z} \|_1$ so that SAV performs similar to BP. We note that
\begin{align*}
(1- p) \, \| \mathbf{z} \|_1 + p \, \| \mathbf{z} -  \textbf{1}_N \|_1
= \sum_{i=1}^N f(z_i),
\end{align*}
where
\begin{align}
\label{eq:ftn_f}
f(t) :=
\begin{cases}
      - t + p & \text{if~} t < 0, \\
      (1-2p) \, t + p & \text{if~} 0 \leq t < 1, \\
      t - p & \text{if~} t \geq 1 .
   \end{cases}
\end{align}

\section{Box-Constrained Sum of Smoothed $\ell_0$}
\label{sec:BSSL0}
L0 and SL0 utilize the $\ell_0$ norm and its smoothed version respectively, however, they do not take into account that $\mathbf{x}$ is binary.
On the other hand, Boxed BP, SN, and SAV utilize the $\ell_1$ norm in one way or another and are specifically adjusted to the binary setting.
A natural question arises: Can we achieve a better recovery rate for binary signals by adjusting L0 and SL0 to the binary setting?

We note that Boxed BP takes into account the binariness of $\mathbf{x}$ by imposing the restriction $\mathbf{x} \in [0,1]^N$. It is straightforward to apply this trick to L0 and SL0, and we will call the resulting algorithms \emph{Boxed L0} and \emph{Boxed SL0} respectively.
Boxed L0 is still NP-hard like L0, but Boxed SL0 shows a clear improvement over SL0 while requiring a similar amount of run time (Fig.~\ref{fig:Lt}). However, the recovery rate of Boxed SL0 is significantly worse than Boxed BP or SN.

In this paper, we aim to adapt the SAV method and the restriction $\mathbf{x} \in [0,1]^N$ to SL0, in order to achieve a better performance.
A straightforward adaptation leads to the following formulation. For $\sigma > 0$ small,
\begin{equation}
\begin{aligned}
& \underset{\mathbf{z} \in [0,1]^N}{\text{min}}
 & &  F_{\sigma}(\mathbf{z}) && \text{subject to} & & \Phi \mathbf{z} = \mathbf{y},  \label{eq:straightforward_adapt}
\end{aligned}
\end{equation}
where
\begin{equation}
\begin{split}
& F_{\sigma}(\mathbf{z}) \triangleq  (1-p) \sum_{i = 1}^N \left (1 - e^{-z_i^2/(2 \sigma^2)} \right ) \\
& \quad \quad \quad + p \sum_{i = 1}^N \left ( 1 - e^{-(z_i-1)^2/(2 \sigma^2)} \right ) \\
& = \sum_{i = 1}^N  \left ( 1 - (1-p) \, e^{-z_i^2/(2 \sigma^2)} - p \, e^{-(z_i-1)^2/(2 \sigma^2)} \right )  \label{eq:SWl0}
\end{split}
\end{equation}
and $p = \mathbb{P}(x_j = 1),~\forall  j \in [N]$.
Note that by (\ref{eq:SL0convergence}), we have
\[ \lim_{\sigma \to 0} F_{\sigma}(\mathbf{z}) =(1 - p) \, \| \mathbf{z} \|_0 + p \, \| \mathbf{z} -  \textbf{1}_N \|_0 \] so that $F_0 (\mathbf{z})$ can be approximated by $F_{\sigma}(\mathbf{z})$ with small $\sigma > 0$.

Next, we will use a weight function to incorporate the restriction $\mathbf{z} \in [0,1]^N$ into the function $F_{\sigma}(\mathbf{z})$. For integers $k \geq 1$, let
\begin{align*}
&\begin{split}
w_{k}(t) & \triangleq \begin{cases}
      1 & \text{if~}  0 \leq t \leq 1 \\
            k &  \text{otherwise}.
	\end{cases}
   	\end{split} \\
 \end{align*}


For $\sigma > 0$ and integers $k \geq 1$, we define
\begin{align*}
&  F_{\sigma, k}^{\text{boxed}}(\mathbf{z}) \\
&\triangleq  \sum_{i = 1}^N  w_{k}(z_i) \left ( 1 - (1-p) \, e^{-z_i^2/(2 \sigma^2)} - p \, e^{-(z_i-1)^2/(2 \sigma^2)} \right ) . \nonumber
\end{align*}
Note that since $1 - (1-p) \, e^{-t^2/(2 \sigma^2)} - p \, e^{-(t-1)^2/(2 \sigma^2)} > 0$ for all $t \in \mathbb{R}$, minimizing $F_{\sigma, k}^{\text{boxed}}(\mathbf{z})$ forces $w_k(z_i)$ to be small so that all $z_i$'s lie within $[0,1]$.
In this way, the restriction $\mathbf{z} \in [0,1]^N$ is incorporated into the cost function.
Our optimization problem now reads as follows:
For $\sigma > 0$ small and $k \in \mathbb{N}$ large,
\begin{equation*}
\begin{aligned}
& \underset{\mathbf{z} \in {\mathbb{R}}^N}{\text{min}}
 & &  F_{\sigma, k}^{\text{boxed}}(\mathbf{z}) && \text{subject to} & & \Phi \mathbf{z} = \mathbf{y} .
\end{aligned}
\end{equation*}
To solve this problem, we propose an algorithm which is based on the gradient descent method and is implemented similarly as algorithms in \cite{Mohimani09,Mohammadi14}.
A major difference in our algorithm is that the cost function $\sum_{i = 1}^N  \big( 1 - \exp \big(\frac{- z_i^2}{2 \sigma^2} \big) \big)$ of SL0 is replaced with $F_{\sigma, k}^{\text{boxed}}(\mathbf{z})$ which is designed specifically for binary signals by adapting the formulation of SAV \cite{Nagahara15}.

	\begin{algorithm}
		\caption{Box-Constrained Sum of Smoothed $\ell_0$ (BSSL0)}
				\begin{algorithmic}[1]
			\State \textbf{Data:} Measurement matrix $\Phi \in \mathbb{R}^{m \times N}$, observation $\mathbf{y} \in \mathbb{R}^m$, probability distribution prior $p = \mathbb{P}(x_j = 1)$.
			\State \textbf{Parameters:} 
            Iters and $L$ are the number of iterations in the outer and inner loops respectively, $\mu$ is a step-size parameter for gradient descent, and $d$ is a decreasing factor for $\sigma$.
			\State \textbf{Initialization:} $\hat{\mathbf{x}}=\Phi^\top(\Phi \Phi^\top)^{-1} \mathbf{y}$, $\sigma = 2 \max |\hat{\mathbf{x}}|$, \\ \quad 
            $k = 1 + N p/\text{Iters}$;
			\For{$1 : \text{Iters}$}
			\For{$1 : L$}
			\State $\hat{\mathbf{x}} \leftarrow \hat{\mathbf{x}} - \sigma^2 \mu \nabla F_{\sigma, k}^{\text{boxed}}(\mathbf{z})$;
\quad \emph{\% gradient descent}
			\State  $\hat{\mathbf{x}} \leftarrow \hat{\mathbf{x}} - \Phi^\top(\Phi \Phi^\top)^{-1}(\Phi \hat{\mathbf{x}} - \mathbf{y})$; \quad \emph{\% projection}
			\EndFor
			\State $\sigma = \sigma \times d$;
            \State$k = k + N p/\text{Iters}$;
			\EndFor
			 \State $\hat{\mathbf{x}} \leftarrow \textbf{round}(\hat{\mathbf{x}})$; \quad     \emph{\% round to a binary vector}
		\end{algorithmic}
		\label{algorighm:BSSL0}
	\end{algorithm}		

The proposed algorithm is comprised of two nested loops. In the outer loop, we slowly decrease $\sigma$ and iteratively search for an optimal solution from a coarse to a fine scale by decreasing $\sigma$ by a factor of $0 < d < 1$. As $\sigma$ decreases, we also gradually increase $k$ so that a larger penalty is put on solutions that have entries outside the range $[0,1]$.
The inner loop performs a gradient descent of $L$ iterations for the function $F_{\sigma, k}^{\text{boxed}}(\mathbf{z})$, where $\sigma$ and $k$ are given from the outer loop. In each iteration of the gradient descent, the solution is projected into the set of feasible solutions $\{ \mathbf{z} : \Phi \mathbf{z} = \mathbf{y} \}$.

Numerical experiments in Section \ref{sec:Experiments} show that for binary signals the proposed algorithm outperforms all other algorithms (BP, Boxed BP, SN, SAV, SL0, and Boxed SL0).

As already mentioned, our algorithm is implemented similarly as SL0 \cite{Mohimani09,Mohammadi14}. The parameters used in our algorithm are exactly the same as in \cite{Mohimani09} except $k$ and $p$. As justified in \cite[Section IV-B]{Mohimani09}, we set the initial estimate of $\mathbf{x}$ as the minimum $\ell_2$ norm solution of $\Phi \mathbf{z} = \mathbf{y}$, i.e., $\hat{\mathbf{x}}=\Phi^\top(\Phi \Phi^\top)^{-1} \mathbf{y}$.
The initialization value for $\sigma$ is discussed in \cite[Remark 5 in Section III]{Mohimani09}. Also, the choice of the step-size $\sigma^2 \mu$ for gradient descent is justified in \cite[Remark 2 in Section III]{Mohimani09} and the choice of $k$ in \cite[Lemma 1]{Mohammadi14}.

The gradient of $F_{\sigma, k}^{\text{boxed}}(\mathbf{z})$ used in Algorithm \ref{algorighm:BSSL0} is given by
\begin{align*}
\begin{split}
 \nabla F_{\sigma, k}^{\text{boxed}}(\mathbf{z}) &= \left( \frac{\partial F_{\sigma, k}^{\text{boxed}}(\mathbf{z})}{\partial z_1}, \dots,  \frac{\partial F_{\sigma, k}^{\text{boxed}}(\mathbf{z})}{\partial z_N} \right)^\top ,
 \end{split}
\end{align*}
where
\begin{align*}
 \frac{\partial F_{\sigma, k}^{\text{boxed}}(\mathbf{z})}{\partial z_i} &= \frac{w_{k}(z_i)}{\sigma^2} \left ((1-p) \, z_i \, e^{-z_i^2/(2 \sigma^2) } \right.   \\
 &  \qquad \left. + \; p \, (z_i -1) \, e^{-(z_i-1)^2/(2 \sigma^2)}  \right )
\quad a.e.
\end{align*}
This is derived using the fact that $w_{k}'(t) = 0$ for all $t$ except $t = 0,1$; we have set $w_{k}'(0) = w_{k}'(1) = 0$ in the implementation.
Let us point out that the discontinuity of $w_{k}(t)$ at $t = 0,1$ does not deteriorate the performance of gradient descent.
One can replace the function $w_{k}(t)$ with a smooth function, however, at the cost of increased run time.

\section{Numerical Experiments} \label{sec:Experiments}

In this section, we compare the performance of our algorithm BSSL0 with other CS/BCS algorithms described in Section \ref{sec:BCS}.
The MATLAB codes for the experiments are available in \cite{myCode}.

\subsection{Experiment 1: Binary Sparse Signal Reconstruction}

In this experiment, we tested BSSL0 with randomly generated binary signals and compared it with other CS/BCS algorithms. 
Random Gaussian matrices are considered for the measurement matrix $\Phi \in \mathbb{R}^{40 \times 100}$, that is, all entries of $\Phi$ are drawn independently from the standard normal distribution.
The parameter $p$ is varied from $0$ to $1$ by step-size $0.05$, and a binary signal $\mathbf{x} \in \{ 0 , 1 \}^{100}$ is generated by drawing its entries independently with $\mathbb{P}(x_i = 1) = p$ and $\mathbb{P}(x_i = 0) = 1 -p$. For $\Phi$ and $\mathbf{x}$, we compute the measurement vector $\mathbf{y} = \Phi \mathbf{x}$ and run the respective algorithms introduced in section II (BP, Boxed BP, SN, SAV, SL0, Boxed SL0, and BSSL0) to obtain a solution vector $\mathbf{z}$ as a approximated reconstruction of $\mathbf{x}$. Additionally, we consider the Orthogonal Matching Pursuit (OMP) \cite{Tropp2007} which is a fast greedy  algorithm for sparse signal reconstruction.
The following are considered for the performance evaluation: (i) \textbf{Failure of Perfect Reconstruction (FPR)}: $0$ if $\mathbf{z} = \mathbf{x}$ (successfully recovered the signal perfectly) and $1$ if $\mathbf{z} \neq \mathbf{x}$ (failed to recover perfectly); (ii) \textbf{Noise Signal Ratio (NSR)}: NSR = $\frac{\| \mathbf{x} - \mathbf{z} \|_2}{\|\mathbf{x}\|_2}$; (iii) \textbf{Run time}.
For each $p$, experiments are repeated $10,000$ times and the results are averaged. For SN, we set the parameter $\lambda$ to be $100$ as fine-tuned in \cite{Wang13}. For BSSL0, we set $\sigma_{\text{min}} = 0.1$, $d = 0.5$, $\mu = 2$, and $L = 1000$.

\begin{figure}[htp]
\centering
\includegraphics[width=1\linewidth]{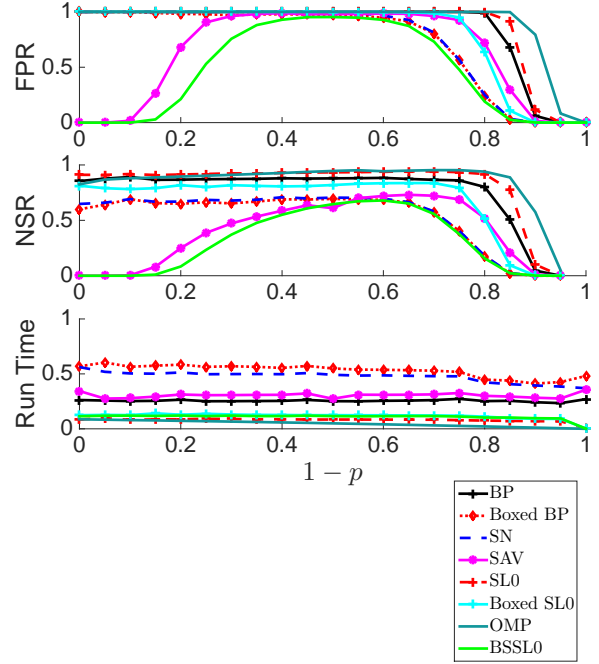}
\caption{Results for Experiment 1.}
\label{fig:Lt}
\end{figure}
In Fig.~\ref{fig:Lt}, BSSL0 shows a better recovery rate than other CS/BCS algorithms and also shows a run time comparable to SL0.

\subsection{Experiment 2: Bitonal Image Reconstruction}

\begin{figure}[H]
\centering
\includegraphics[width=0.48\linewidth]{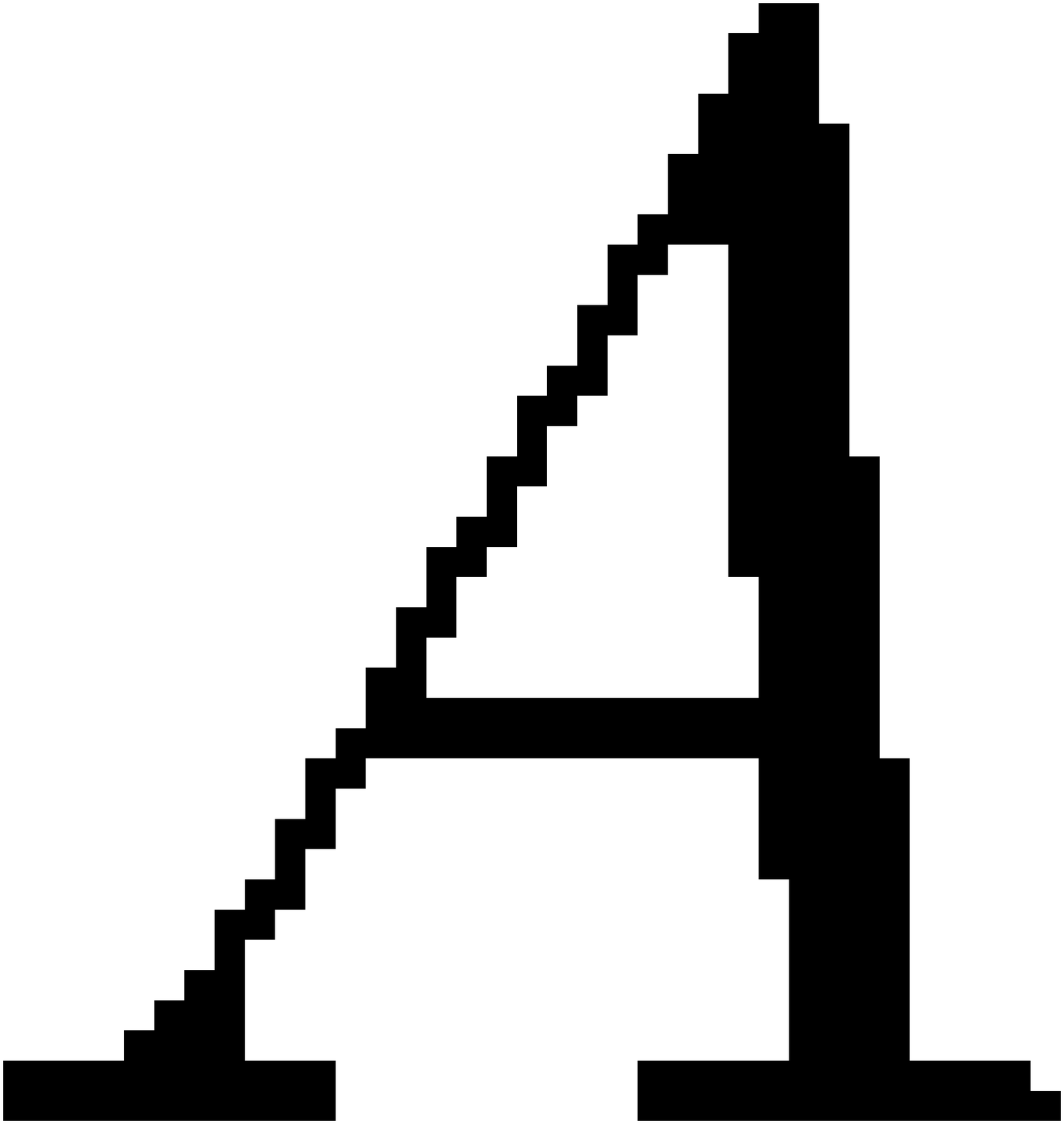}~
\includegraphics[width=0.48\linewidth]{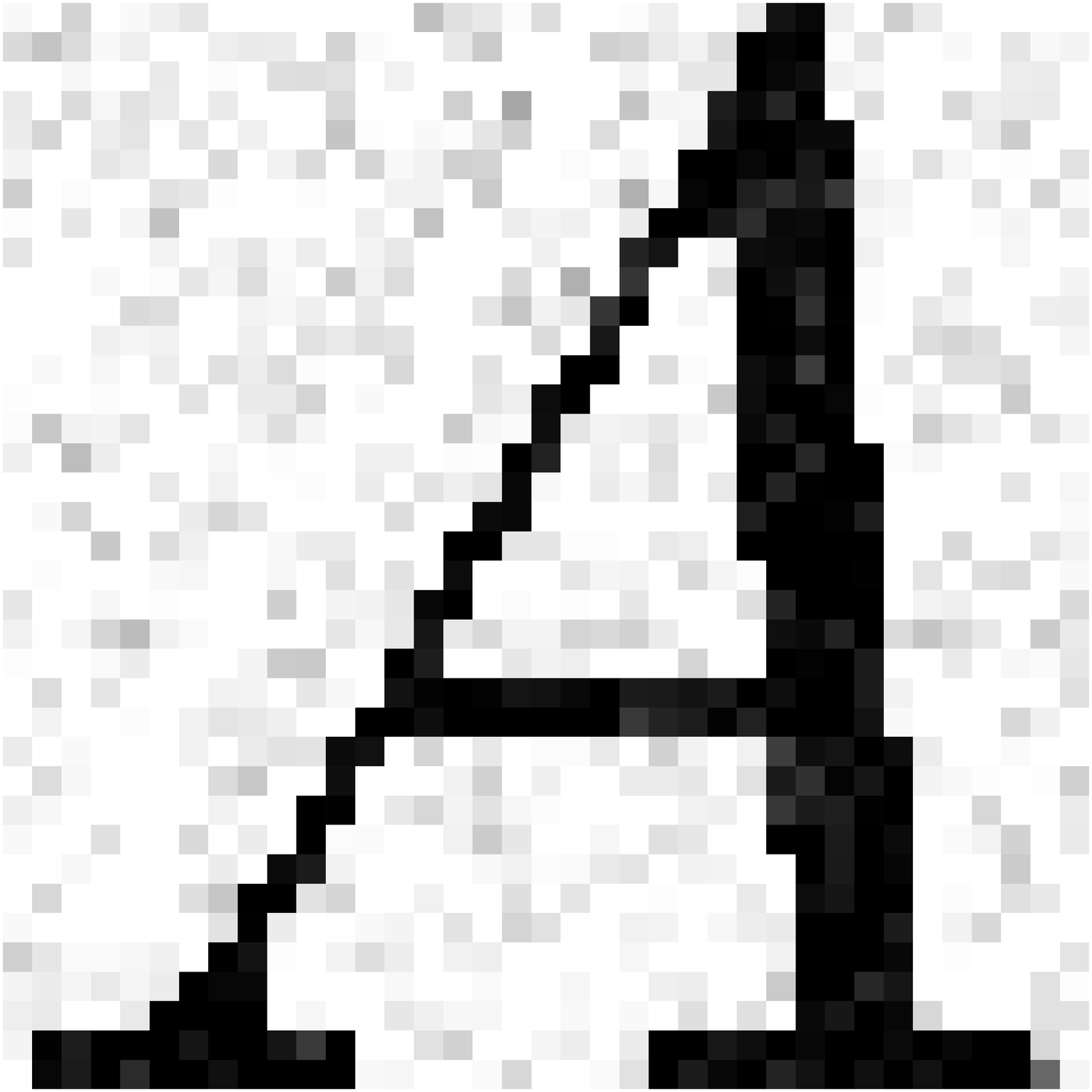}
\caption{Original image (left) and the image corrupted by Gaussian noise (right).}
\label{fig:img_orig}
\end{figure}

As in \cite{Nagahara15}, we considered reconstruction of the $37\times 37$-pixel bitonal image given in Fig.~\ref{fig:img_orig} (left).
Following the same setup in \cite{Nagahara15}, we added to each pixel a random Gaussian noise with mean-zero and standard deviation of $0.1$, as shown in Fig.~\ref{fig:img_orig} (right).

\begin{figure}[H]
\centering
\includegraphics[width=0.48\linewidth]{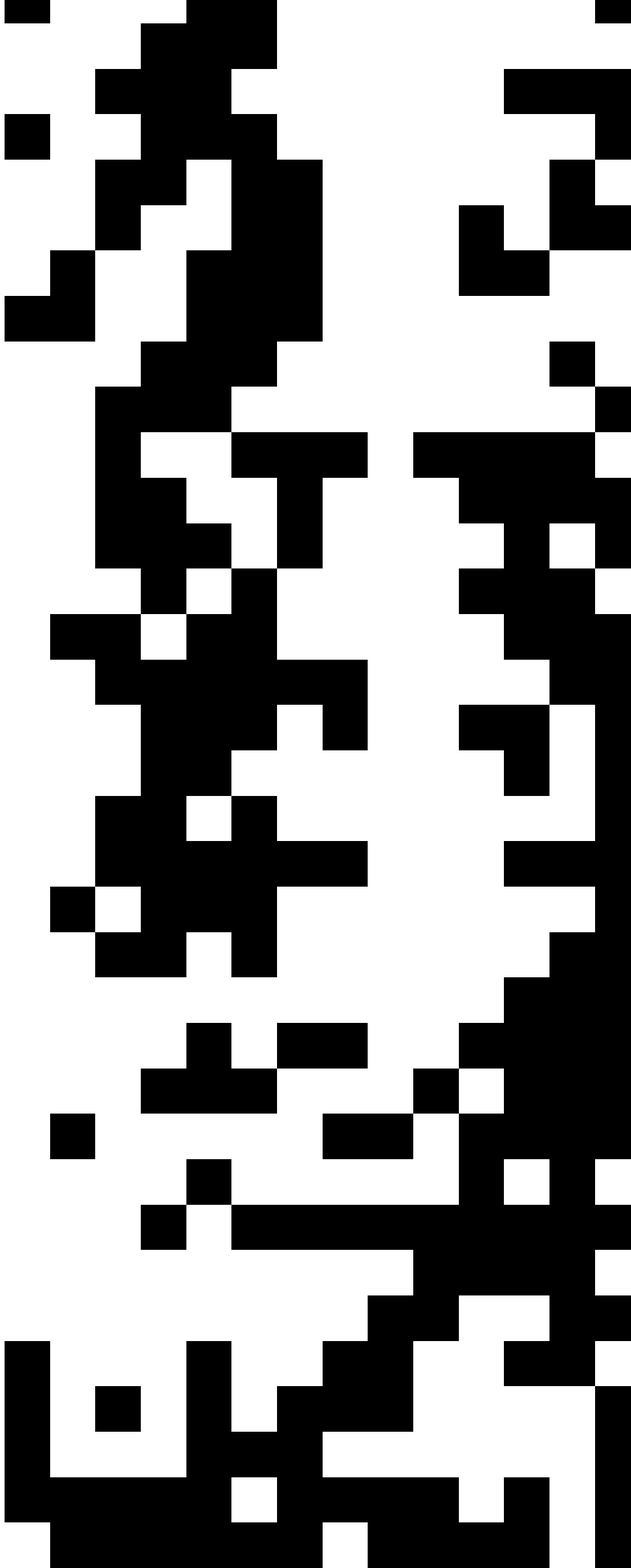}~
\includegraphics[width=0.48\linewidth]{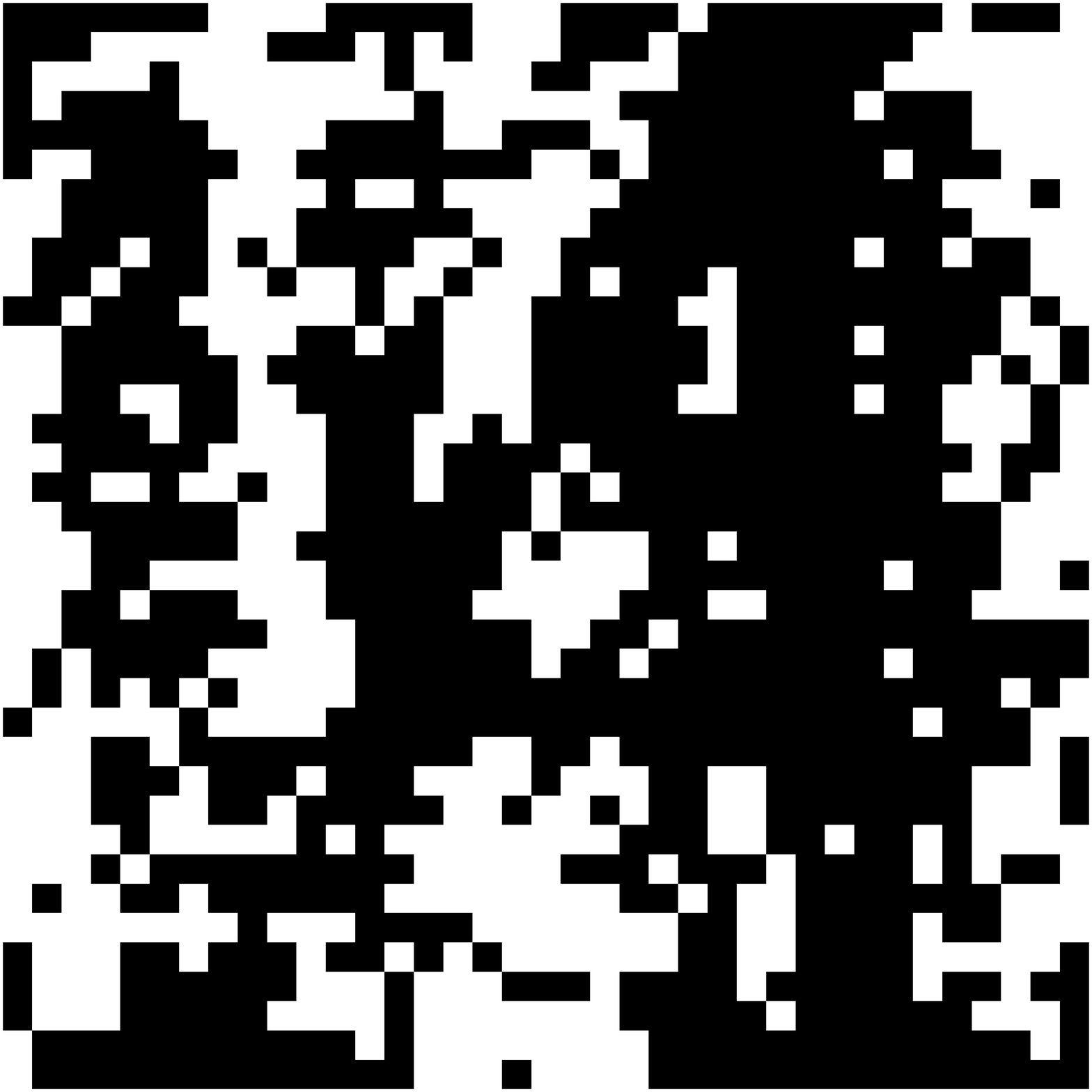}~ \\
\includegraphics[width=0.48\linewidth]{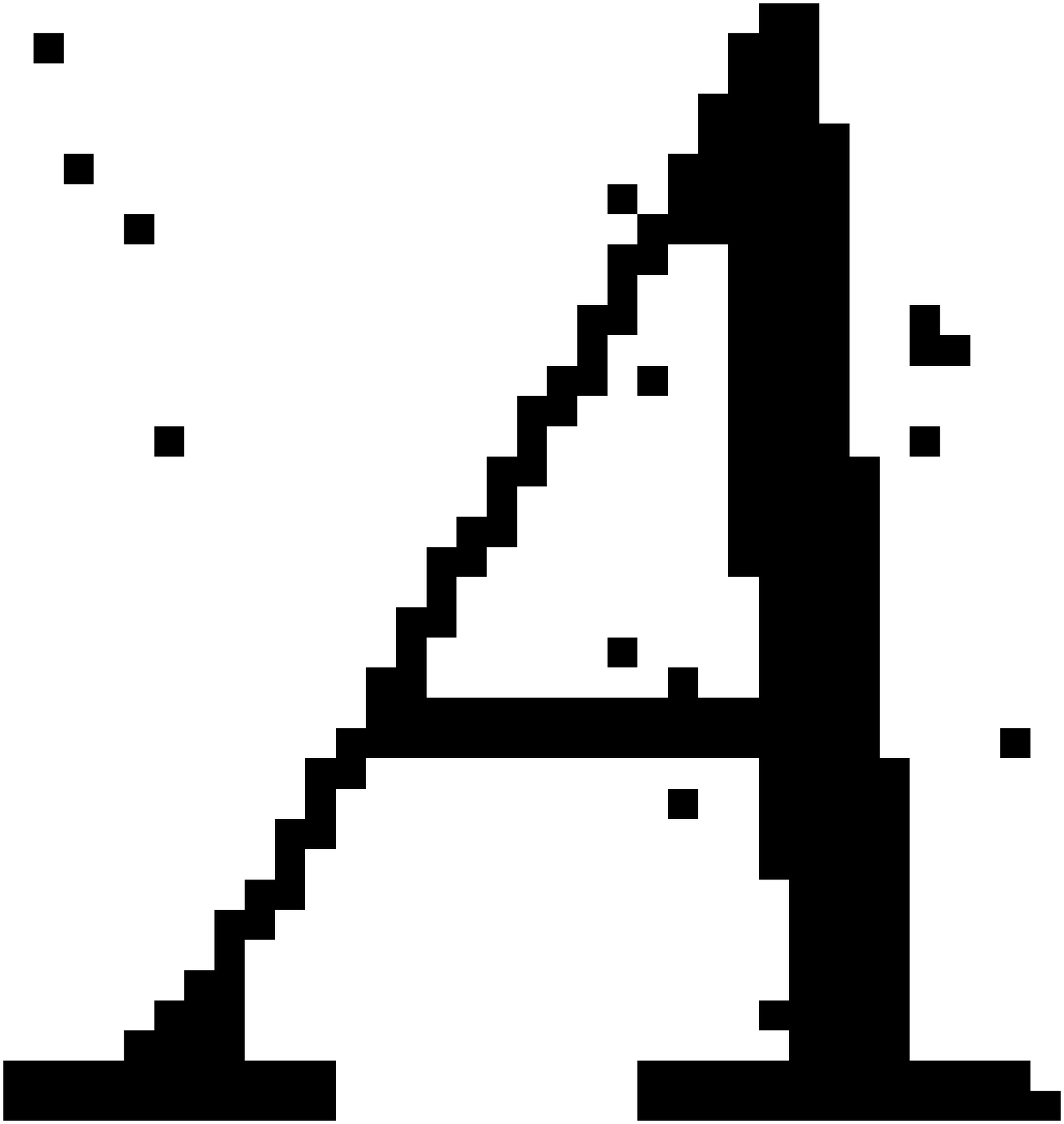}~
\includegraphics[width=0.48\linewidth]{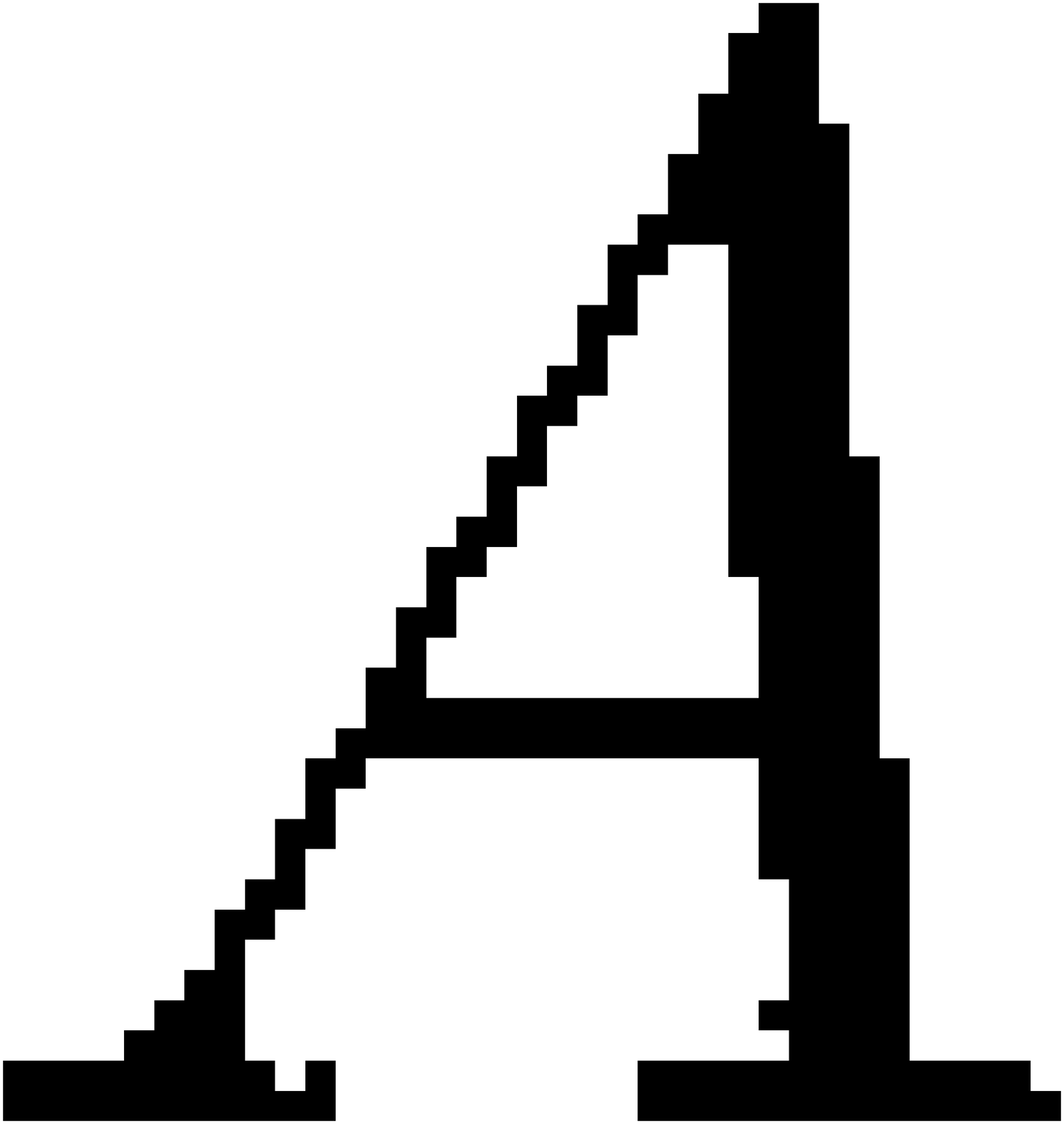}
\caption{Reconstructed images by BP (upper left), SN (upper right), SAV (lower left), and the proposed method BSSL0 (lower right).}
\label{fig:reconstruction}
\end{figure}

\noindent
The noisy image is represented by a real-valued $37 \times 37$ matrix $X$
and we apply the discrete Fourier transform (DFT) to obtain
\[
\hat{X} = WXW \;\;  \in \mathbb{C}^{37 \times 37} ,
\]
equivalently,
\[
\mathrm{vec}(\hat{X}) = (W\otimes W) \, \mathrm{vec}(X)  \;\;  \in \mathbb{C}^{1369} ,
\]
where $W = [ \omega^{k, \ell} ]_{k,\ell =0}^{K-1}$ with $K=37$ and $\omega = e^{-2 \pi i / K}$ is the $K$-point DFT matrix.
As in \cite{Nagahara15}, we randomly subsampled $\mathrm{vec}(\hat{X}) \in \mathbb{C}^{1369}$ to obtain a half-sized vector $\mathbf{y} \in \mathbb{C}^{685}$ and set the measurement matrix $\Phi$ as the corresponding $685 \times 1369$ submatrix of $W\otimes W$.
Fig.~\ref{fig:reconstruction} shows the reconstructed images by BP, SN, SAV, and BSSL0, all with entrywise rounding off to $\{ 0 , 1 \}$.
For SN, an optimal tuning parameter $\lambda$ was searched from $50$ to $1000$ by stepsize $50$ and the value $\lambda = 800$ was chosen. For SAV and BSSL0, as in \cite{Nagahara15}, we chose the parameter $p = \mathbb{P}(x_j = 0) = 0.5$ as a rough estimate for the sparsity of the bitonal image (see \cite{Nagahara15}).
We set $\sigma_\text{min} = 0.01$, $d =  0.9$, $\mu = 2$, and $L = 3$ for the parameters of BSSL0.
The respective run time for BP, SN, SAV, and BSSL0 are also given in Tab.~\ref{tab:timesonsumption}.

\begin{table}[H]
\caption{The Run Time Comparison}
\begin{center}
\label{tab:timesonsumption}
    \begin{tabular}{ | l | l | l | l | }
    \hline
    \text{Algorithm} & Run Time\\ \hline
    Basis Pursuit &  185.2044 seconds  \\ \hline
    SN      &  406.1007 seconds \\ \hline
    SAV & 191.5366 seconds \\ \hline
    \textbf{BSSL0} (proposed) & \textbf{0.92577 seconds}  \\
    \hline
    \end{tabular}
\end{center}
\end{table}


\section{Conclusion}
\label{sec:conclusion}

In this work, we proposed a fast algorithm (BSSL0) for reconstruction of binary signals which is based on the gradient descent method and smooth relaxation techniques. We showed that for binary signals our algorithm outperforms other CS/BCS methods in terms of the recovery rate and speed. Future work includes a detailed analysis of BSSL0 in stability/robustness and extensions to ternary and finite alphabet signals.

\section*{Acknowledgment}
T.~Liu and D.~G.~Lee acknowledge the support of the DFG Grant PF 450/6-1. The authors are grateful to Robert Fischer and G\"otz E.~Pfander for their helpful suggestions. The authors thank anonymous reviewers for their comments.



%
\IEEEpeerreviewmaketitle



%
\bibliographystyle{IEEEtran}
\bibliography{IEEEabrv,BSSL0}

\end{document}